\documentstyle[sprocl,epsf]{article}

\newcommand{\vev}[1]{\langle #1 \rangle}

\begin{document}

\title{
CP VIOLATION VIA $\rho \omega$ INTERFERENCE \footnote{
Talk presented by R. Enomoto at "International Conference on 
B physics and CP violation", Mar. 1997
, Hawaii, USA.}
}

\author{ RYOJI ENOMOTO }

\address{High Energy Accelerator Research Organization, KEK,
Ibaraki 305, Japan}

\author{ MASAHARU TANABASHI }

\address{Department of Physics, Tohoku University,
Sendai 980-77, Japan}


\maketitle\abstracts{
We consider $B^{\pm,0}\rightarrow \rho^0(\omega)h^{\pm,0}$, where
$\rho^0(\omega)$ decays to $\pi^+\pi^-$ and $h$ is any hadronic final state,
such as $\pi$ or $K$. We find a large direct $CP$ asymmetry in $B$-meson decays
via $\rho \omega$ interference.
A possible method to determine weak phases, such as $\phi_{2,3}$, is 
discussed.
The experimental feasibility is also shown.
}

\section{Introduction}

The standard model of $CP$ violation \cite{km} predicts large $CP$
asymmetries in the $B$-meson system.\cite{sanda}
Many experimental attempts to detect $CP$ violations of the $B$ meson will
be carried out towards the next century.\cite{belle,babar,herab}

Unlike $CP$ violations in the neutral $B$ meson, the $CP$
asymmetry of the charged $B$ meson can be caused solely by
direct $CP$ violations, which only occur through 
interference between two amplitudes having different weak and strong
phases. 
In the standard model a weak-phase difference is provided by a
different complex phase of the Kobayashi-Maskawa (KM) matrix
elements \cite{km} of the tree and penguin diagrams, while the strong phase
is given by the absorptive parts of the corresponding diagrams. 

Nonperturbative resonance states, in which we know the behavior of 
the large absorptive part by using the Breit-Wigner shape, are ideal places 
to obtain large but controllable $CP$ asymmetry.
The $CP$ violations via radiative decays of the $B$ meson were predicted
by Atwood and Soni.\cite{gammak,gammaa}
The role of charmonium resonances in the $CP$ violation of $B^\pm$ decay
has been discussed by Eilam, Gronau and Mendel.\cite{eilam} 
Lipkin discussed the use of $\rho$-$\omega$ interference as a trigger
of direct $CP$ violation in neutral $B$-meson decay using
a simple quark-model analysis.\cite{kn:lipkin}

In this talk we present a systematic analysis of the $CP$ asymmetries in
$B^{\pm, 0}\rightarrow \rho^0 (\omega) h^{\pm,0} \rightarrow
\pi^+\pi^- h^{\pm,0}$  
via $\rho \omega$ interference,\cite{nambu} 
where $h$ is any hadronic final state, such as $\pi$, $\rho$, $K$,
or $K^*$.
We find large $CP$ asymmetries at the interference region.\cite{et}

\section{Mechanism}

\begin{figure}
\epsfbox{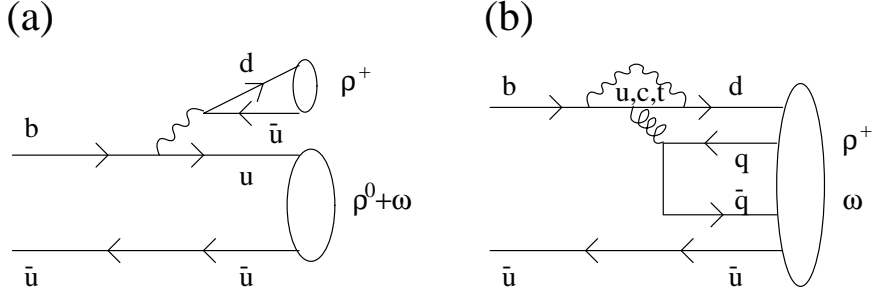}
\vskip 1cm
\caption{Examples of the Feynman diagrams of the decay
$B^-\rightarrow \rho^0 (\omega) \rho^-$;
(a) tree and (b) penguin diagram.}
\label{feynman}
\end{figure}

Figures \ref{feynman} (a) and (b) are examples of quark-level diagrams
of the tree (a) and penguin (b) amplitudes for 
$B^{-}\rightarrow \rho^0 (\omega) \rho^{-}$.
Considering the quark components, diagram (a) gives the final
state of $\rho^0+\omega$, diagram (b) contributes solely to
the $\omega$ meson. 
The standard model predicts a weak phase difference, 
$\phi_2 = {\rm arg}((V_{ud}V^*_{ub})/(V_{td}V^*_{tb}))$.
The absorptive part (strong phase) is provided by both the
$\rho$-$\omega$ interference and the Bander-Silverman-Soni
mechanism (the quark loop absorptive part in the penguin
diagram).\cite{kn:BSS}

The size of the interference effect can be evaluated as
\begin{equation}
\frac{\vev{\pi^+ \pi^-|J_0^\mu|0}\epsilon_\mu}
     {\vev{\pi^+ \pi^-|J_3^\mu|0}\epsilon_\mu}
  \simeq
        \frac{g_\omega}{g_\rho} 
        \frac{g_\rho^2}{m_\omega^2 - i\Gamma_\omega m_\omega -s}
        \left[
           \frac{g_\omega}{3} \frac{e^2}{-s} g_\rho + g_{\rho\omega}
        \right],
\label{eq:lab0}
\end{equation}
where $s$ denotes the invariant mass square of $\pi^+\pi^-$, and
$g_\omega$, $g_\rho$, and $g_{\rho\omega}$
 are the decay constants of the $\omega$ and $\rho$ mesons,
and $\rho$-$\omega$ mixing amplitude,
respectively.
Considering
\begin{equation}
\frac{\Gamma(\omega\rightarrow \pi^+ \pi^-)}
     {\Gamma(\rho  \rightarrow \pi^+ \pi^-)}
 = \frac{1}{(m_\rho^2-m_\omega^2)^2+\Gamma_\rho^2 m_\rho^2}
        \left[
           \frac{g_\omega}{3} \frac{e^2}{-m_\omega^2} g_\rho 
          + g_{\rho\omega}
        \right]^2,
\label{eq:lab4}
\end{equation}
and
$\Gamma (\omega\rightarrow \pi^+\pi^-)=0.19$MeV,
$\Gamma (\rho\rightarrow \pi^+\pi^-)=\Gamma_{\rho}=150$MeV,
and 
$\Gamma_{\omega}=8.4$MeV 
into Eq.(\ref{eq:lab4}), we find
\begin{equation}
  \frac{g_\omega}{3} \frac{e^2}{-m_\omega^2} g_\rho + g_{\rho\omega}
  \simeq 0.63 \Gamma_\omega m_\omega,
\label{eq:lab5}
\end{equation}
where the sign is determined from $e^+ e^- \rightarrow \pi^+ \pi^-$ 
near to the $\rho$-meson mass.

In the ideal case, 
we obtain the maximum strong-phase difference at $s=m_\omega$.

\section{Determination of Weak Phases}

We next discuss how we can extract the weak phase from
the observed $CP$ asymmetries.

The relevant hadronic form factors are parametrized by six parameters. 
Including one weak phase difference, the amplitudes are written in
terms of seven unknown parameters.
On the other hand, we will measure branching ratios of
$B^- \rightarrow \rho^0 h^-$, $B^+ \rightarrow \rho^0 h^+$,
$B^- \rightarrow \omega h^-$, and $B^+ \rightarrow \omega h^+$.
Also, CP conserving $\rho \omega$ interference, e.g., the pole
position and the interference amplitude will be measured to fix two
parameters. 
Finally, the CP asymmetry in the $\rho \omega$ interference region
gives two types of information. 
Setting aside one normalization factor, we should experimentally
have seven measurements. We can therefore determine such weak phases
as $\phi_2$ and $\phi_3$.
We emphasize here that this method does not rely on the factorization 
assumption.
We also note that the existence of the electroweak penguin operators
does not affect this determination.

\section{Estimation}

The hadronic form factors are evaluated in Ref \cite{et} so as to
demonstrate the feasibility to detect $CP$ asymmetries in this mode.
We have taken into account the 
finite radiative correction to the penguin-type on-shell quark
amplitude coming from the tree
hamiltonian.\cite{kn:Desh,kn:fleischer,kn:kramer}
The factorization was assumed with $N_c$ being treated as a parameter
to parametrize uncertainty of this assumption.
Measurements of the branching fractions of $B\rightarrow D$
decays indicate
$N_c\simeq 2\sim 3$ in this parametrization.\cite{kn:Rod96}
In the reference,\cite{et} we used $N_c=2$ and $N_c=\infty$, and 
the gluon momentum
in the Penguin diagram ($k^2$) to be $0.5m_b^2$ and $0.3m_b^2$.
For the meson form factors, we used the BSW model.\cite{bsw}

A summary is listed in Table \ref{table5} for the $N_c=2$ 
and $k^2=0.5m_b^2$ case.
\begin{table}
\begin{tabular}{l|rrrrrrr}
\hline\hline
Mode & BR & $A(\rho^0)$ & $A(\omega)$ & $A(\rho \omega)$
& $A^{max}(\rho \omega)$ & $A^0(\rho \omega)$ & $N(B\bar{B})$\\
 & $\times 10^{-8}$ & \% & \% & \% & \% & \% & $\times10^8$\\
\hline
$\rho^0 \rho^-$ &
2100 & 0 & 10 & 13 & 26 & 11 & 70\\
$ \rho^0 K^{*-}$ &
720 & -36 & -19 & -45 & -79 & -19 & 12\\
$ \rho^0 \pi^-$ &
660 & -6 & 11 & 16 & 37 & 18 & 29\\
$ \rho^0 K^-$ &
62 & -41 & -26 & -82 & -91 & -67 & 7.6
\\\hline\hline
\end{tabular}
\caption{
Asymmetries for the various decay modes:
BR is the branching ratios in unit of $10^{-8}$;
$A(\rho^0)$ and $A(\omega)$ are asymmetries for
$B^-\rightarrow \rho^0 h$ and $\omega h$ modes, respectively.
$A(\rho \omega)$ is that in the region of $M(\omega)\pm
\Gamma(\omega)$. $A^{max}(\rho \omega)$ is the maximum
asymmetry in this region. $A^0(\rho \omega)$ is that under the assumption
of ``zero hadronic phase".
$N(B\bar{B})$ is the necessary number of $B\bar{B}$ events
in order to obtain a 3$\sigma$ asymmetry in $\rho-\omega$ interference
region.
}
\label{table5}
\end{table}
$A(\rho^0)$ and $A(\omega)$ are the asymmetries
of the $\rho^0 h$ and $\omega  h$ modes.
$A(\rho \omega)$ is the mean asymmetry of the $M(\pi^+\pi^-)$
invariant mass spectra around $M(\omega)\pm\Gamma(\omega)$,
and $A^{max}(\rho \omega)$ is the maximum asymmetry in this
region. $A^0(\rho \omega)$ is obtained by assuming the
``zero hadronic phase".
The branching ratios were also estimated using this formalism.
The $CP$ asymmetries via $\rho$-$\omega$ interference are large
($>$ 10\%) in most cases.

\section{Experimental Feasibility}
\begin{figure}
\epsfysize 14cm
\epsfbox{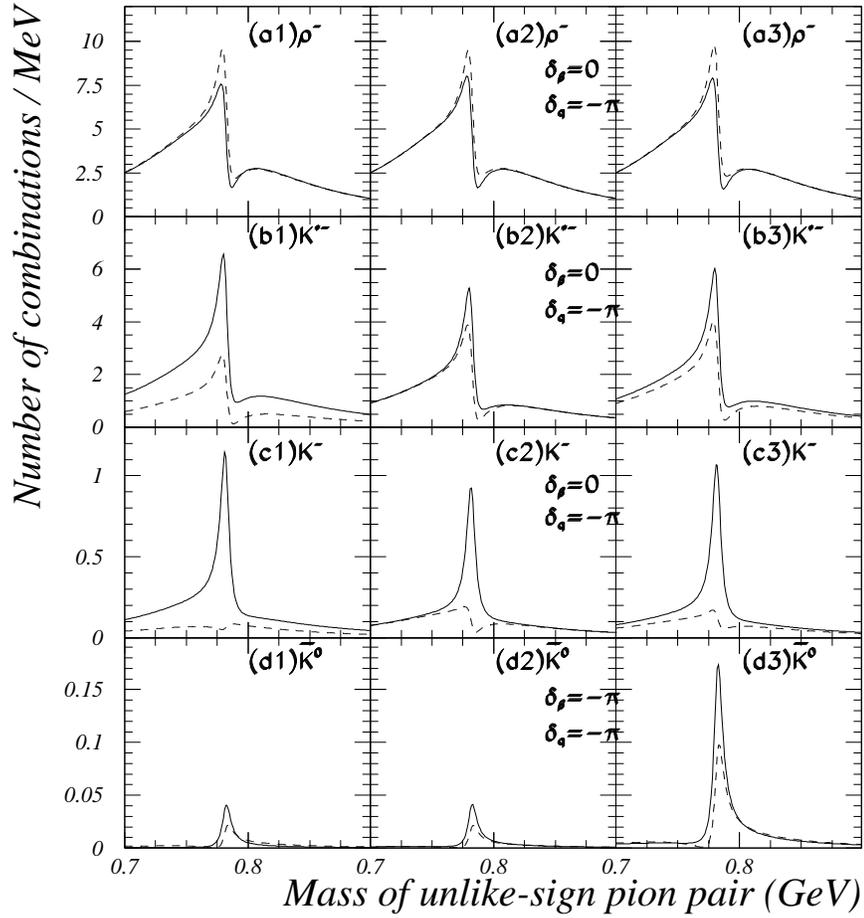}
\caption{
Expected invariant mass spectra of unlike-sign pion pairs.
The solid lines are for $B^+$ or $B^0$ decays,
and the dashed ones are for $B^-$ or $\bar{B^0}$
decays.
The vertical scale is the differential yield for
$(\pi^+\pi^-)h^{\pm ,0}$
combinations, and is normalized to give
the number of entries at $10^8~B\bar{B}$ events, assuming an
100-\% acceptance.
The details concerning the notations (a1)-(d3) are described in the text.
}
\label{pattern}
\end{figure}

We demonstrate the asymmetry patterns ($\pi^+\pi^-$ invariant
mass spectra) in Figures \ref{pattern} (a1)-(d3),
where (a), (b), (c), and (d) denote the $B^-\rightarrow \rho^0\rho^+$,
$\rho^0 K^{*-}$, $\rho^0 K^-$, and $B^0\rightarrow \rho^0 \bar{K}^0$
decay modes,
respectively.
The results using ($N_c$, $k^2$) = (2, $0.5m_b^2$)  are given in
Figures \ref{pattern} (a1), (b1), (c1), and (d1), and
those with ($N_c$, $k^2$) = (2, $0.5m_b^2$) 
are shown in Figures \ref{pattern} (a3), (b3),
(c3), and (d3).
In Figures \ref{pattern} (a2), (b2), (c2), and (d2), we assumed ``zero
short distance hadronic phases" with ($N_c$, $k^2$) = (2, $0.5m_b^2$) .
The solid lines are for $B^+$ or $B$ and the dashed ones are
for $B^-$ or $\bar{B}$. 
Here, we have assumed the KM matrix of the Wolfenstein parametrization
($\lambda=0.221$, $\rho=-0.12$, $\eta=0.34$ and $A=0.84$), which
corresponds to  
the $(\phi_1,\phi_2,\phi_3)=(15,55,110)$ degrees.
The branching ratios in these parameters are given in Table \ref{table5}.
The vertical scales are normalized to give the
number of entries at $10^8$ $B\bar{B}$ events with an 100-\% acceptance.
Drastic asymmetries appear around the $\omega$ mass region.

In order to check the feasibility for detecting this $CP$
asymmetry,
we performed a simulation assuming the BELLE detector of 
the KEK B-factory,\cite{belle} an asymmetric $e^+e^-$ collider
(8 x 3.5GeV). 
The invariant mass 
resolution of $\pi^+\pi^-$ around $\omega$ mass is expected to be
3.2 MeV for the $B\rightarrow \omega h$, $\omega\rightarrow \pi^+\pi^-$
decay; this is enough to resolve the interference pattern.
Here, the momentum resolution is derived from
$(dP_T/P_T)^2=(0.001P_T/1{\rm GeV})^2+0.002^2$. 
In the case of a symmetric collider, the mass resolution will be
better.
In the case of a hadron machine, the average $B$ mesons' $P_T$ would be
several GeV or more. 
Although the mass resolution slightly deteriorates,
the statistics are sufficient in hadron machines.

In order to suppress
the large background from continuum events under $\Upsilon (4S)$,
we used two cuts in analyzing the $\rho^0 h^{\pm,0}$ decay:\cite{cleo} 
one was that the 
absolute value of the cosine
of the angle between the thrust axes of B decay products 
and the other particles
at center-of-mass-system of $\Upsilon (4S)$ be less than $0.6$;
the other was that the energy of the $B$ candidate be between 
$5.25$ and $5.325$ GeV. 
The beam-energy constraint mass spectra were used.

The results of a simulation for the $B^{\pm}\rightarrow \rho^0h^{\pm}$
($h^{\pm}=\pi^\pm,\rho^\pm,K^\pm,K^{*\pm}$) decay modes are 
summarized in $N(B\bar B)$ of Table \ref{table5}, 
the necessary number of $B\bar B$ events for detecting the 3$\sigma$ $CP$
asymmetry at the $\rho$-$\omega$ interference region. 
The branching ratios quoted in Table \ref{table5} are assumed.
In some of these decay modes, 3$\sigma$-CP violations
are detectable with $10^9$ $B\bar{B}$ events
by the $\rho-\omega$ interference modes.
If a good method can be found to suppress the background from the continuum,
the necessary luminosity can be significantly reduced.
Also, at this conference, CLEO indicated that the branching ratio of
$B\rightarrow \omega K$ is on the 
order of $10^{-5}$. If this is correct, the
necessary luminosity will be greatly reduced to $\sim$30 $fb^{-1}$,
i.e., definitely within an experiment involving a few years.

\section{Conclusion}

We have studied the effect of $\rho$-$\omega$ interference in the
decay modes $B\rightarrow \rho^0 (\omega) h$, 
$\rho^0 (\omega) \rightarrow \pi^+\pi^-$,
where $h$ is any hadronic final state, such as
$\pi$, $\rho$, or $K$.
Although the isospin-violating decay of $\omega\rightarrow\pi^+\pi^-$ 
is a small effect with BR=$2.2$\%, 
the interference at the kinematical region $M(\pi^+\pi^-)\sim
M(\omega)\pm \Gamma_{\omega}$ is enhanced by the $\omega$ pole.
We have shown the $CP$ asymmetry to be 
sufficiently large to be detected.
The $CP$ asymmetry appears in the deformation of the
Breit-Wigner shape of the $\rho^0\rightarrow \pi^+\pi^-$
invariant mass spectrum. 
The prediction of the $CP$ asymmetry is not very sensitive to the
hadronic phase calculation, i.e., a ``sure" prediction.
Any B-factory, even if it is a symmetric $e^+e^-$ collider
or hadron machine, can carry out this measurement.
We only need to accumulate enough statistics
and to have a mass
resolution [$\Delta M(\pi^+\pi^-)$] better than
the width of the $\omega$ meson (8.4MeV) at around the $\omega$ mass region.

\section*{Acknowledgments}
We thank Drs. M. Kobayashi, A. I. Sanda, A. Soni, M. Tanaka and I. Dunietz
for useful discussions. 
We also thank the Belle collaboration for providing the detector simulation
programs.
This work was partially supported by the Inoue Foundation for Science
and the Grant-in-Aid of Monbusho (the Japanese Ministry of Education,
Science, Sports and Culture) \#09740185 and \#09246203.

\end{document}